\documentclass[iop]{emulateapj}

\newcommand{\se}[1]{\mbox{\S\ \ref{sec:#1}}}
\newcommand{\Se}[1]{\mbox{\S\ \ref{sec:#1}}}
\newcommand{\eq}[1]{equation\ (\ref{eq:#1})}
\newcommand{\eqp}[1]{\mbox{eq.\ [\ref{eq:#1}]}}

\newcommand{\eqps}[2]{Eqs.\ [\ref{eq:#1}] and [\ref{eq:#2}]}
\newcommand{\eqss}[3]{equations\ (\ref{eq:#1}), (\ref{eq:#2}), and (\ref{eq:#3})}
\newcommand{\Eq}[1]{Equation\ (\ref{eq:#1})}

\newcommand{\fg}[1]{\mbox{Fig.\ \ref{fig:#1}}}

\newcommand{\Fg}[1]{\mbox{Figure\ \ref{fig:#1}}}

\newcommand{\vs}{vs.}
\newcommand{\ie}{i.e.,}
\newcommand{\eg}{e.g.,}
\newcommand{\cf}{cf.}

\shorttitle{A new condition for the transition from runaway to oligarchic growth}
\shortauthors{Ormel, Dullemond, \& Spaans}

\begin{document}

\title{A new condition for the transition from runaway to oligarchic growth}

\author{C. W. Ormel}
\affil{Max-Planck-Institute for Astronomy, K\"onigstuhl 17, 69117 Heidelberg, Germany}
\affil{Astronomisches Rechen-Institut, Zentrum f\"ur Astronomie der Universit\"at Heidelberg, M\"onchhofstr.\ 12-14, 69120 Heidelberg, Germany}
\email{ormel@mpia.de}

\author{C. P. Dullemond}
\affil{Max-Planck-Institute for Astronomy, K\"onigstuhl 17, 69117 Heidelberg, Germany}
\email{dullemon@mpia.de}

\and

\author{M. Spaans}
\affil{Kapteyn Astronomical Institute, P.O. Box 800, 9700 AV, Groningen, The Netherlands}
\email{spaans@astro.rug.nl}


%\altaffiltext{1}{Visiting Astronomer, Cerro Tololo Inter-American Observatory.  CTIO is operated by AURA, Inc.\ under contract to the National Science Foundation.}
%\altaffiltext{2}{Society of Fellows, Harvard University.}
%\altaffiltext{3}{present address: Center for Astrophysics, 60 Garden Street, Cambridge, MA 02138}
%\altaffiltext{4}{Visiting Programmer, Space Telescope Science Institute}
%\altaffiltext{5}{Patron, Alonso's Bar and Grill}


\begin{abstract}
  Accretion among macroscopic bodies of $\sim$km size or larger is enhanced significantly due to gravitational focusing.  Two regimes can be distinguished.  Initially, the system experiences runaway growth, in which the gravitational focusing factors increase, and bodies at the high-mass tail of the distribution grow fastest.  However, at some point the runaway body dynamically heats its environment, gravitational focusing factors decrease, and runaway growth passes into oligarchic growth.  Based on the results of recent simulations, we reconsider the runaway growth-oligarchy transition.  In contrast to oligarchy, we find that runaway growth cannot be approximated with a two component model (of small and large bodies) and that the criterion of \citet{IdaMakino1993}, which is frequently adopted as the start of oligarchy, is not a sufficient condition to signify the transition.  Instead, we propose a new criterion based on timescale arguments.  We then find a larger value for the runaway growth-oligarchy transition: from several hundreds of km in the inner disk regions up to $\sim$10$^3$ km for the outer disk.  These findings are consistent with the view that runaway growth has been responsible for the size distribution of the present day Kuiper belt objects.  Our finding furthermore outline the proper initial conditions at the start of the oligarchy stage. 
\end{abstract}





\section{Introduction}
In planet formation theory, the process through which $\sim$km or larger bodies are formed is still not well understood, since a straightforward formation mechanism is lacking \citep{BlumWurm2008,ChiangYoudin2009}.  However, once these bodies appear gravity takes over the accretion process, fulfilling both the role as a sticking agent as well as an important accelerator for growth: the gravitational cross section between two bodies can become much larger than their geometrical cross section. This phenomenon is better known as gravitational focusing and the gravitational enhancement factor in this regime is given by $\sim$$(v_\mathrm{esc}/v)^2$, where $v_\mathrm{esc}$ is the (mutual) escape velocity and $v$ the velocity dispersion of bodies.  Initially, gravitational focusing factors for the biggest bodies increase rapidly, and the system experiences \textit{runaway growth}: $v_\mathrm{esc}$ increases much faster than $v$.  However, at a certain point the stirring capabilities of the biggest body will cause $v$ to increase rapidly.  Gravitational focusing factors decrease and the growth becomes self-regulated since the stirring rate of small bodies is determined by the same big body that accretes them.  Runaway growth has passed into oligarchy \citep{KokuboIda1998}.

\citet{IdaMakino1993} have argued that the runaway growth-oligarchy transition takes place at the point where the stirring power of big bodies first exceeds that of the small bodies, \ie\
\begin{equation}
  2\Sigma_M M > \Sigma_m m,
  \label{eq:IM93}
\end{equation}
where $\Sigma_M$ is the surface density of big bodies of mass $M$ and $\Sigma_m$ that of the small bodies.  \Eq{IM93} can be transformed into a radius, $R_\mathrm{rg/oli}$, indicating the turnover from runaway growth into oligarchy (see below, \eqp{Roli}).  Many works have adopted \eq{IM93} as the start of their oligarchic calculations \citep[\eg][]{ThommesEtal2003,IdaLin2004,Chambers2006,Chambers2008,FortierEtal2007,BruniniBenvenuto2008,MiguelBrunini2008,MordasiniEtal2009}.

In this letter, we will refine the criterion of \citet{IdaMakino1993} and present a new expression for $R_\mathrm{rg/oli}$ (\eqp{R-rg}).  In runaway growth the column density spectrum evolves into a power law, $N(m) = (1/m) d\Sigma/dm \propto m^{-p}$, where $p\approx-2.5$ \citep[][\fg{fig1}a]{WetherillStewart1993,KokuboIda1996,KokuboIda2000,BarnesEtal2009}.  However, for a $p>-3$ index, the stirring power lies at the high-mass end of the population; that is, during runaway growth the stirring power is already moving away from the initial mass ($m_0$).  Despite the stirring, the system continues in its runaway (fast) growth mode.  The point is that \eq{IM93} implicitly assumes that stirring occurs fast and outpaces the accretion, which is true for oligarchy but not for runaway growth. In other words, \eq{IM93} is a necessary condition for oligarchy but not a sufficient one.  Instead, we will argue that the condition for the start of oligarchy is met when the stirring timescale in the two component approximation ($T_\mathrm{vs}$, see below) drops below the accretion timescale that characterizes the runaway regime, $T_\mathrm{rg}$.

In \se{def} we first introduce key definitions and obtain the stirring timescale $T_\mathrm{vs}$ and the accretion timescale $T_\mathrm{ac}$ for a two component system.  In \se{Trg} we present the results of our runaway growth simulations in terms of $T_\mathrm{rg}$, the accretion timescale during runaway growth, which follows from our simulations. \Se{rgot} then presents the new transition radius $R_\mathrm{tr}$ by equating the timescale expressions.  We discusses a few implications and summarize in \se{summ}.

\section{Key definitions and timescales}
\label{sec:def}
The Hill radius $R_h$ and Hill velocity $v_h$ of a single body of mass $M$ and radius $R$ are given by
\begin{equation}
  R_h = a \left( \frac{M}{3M_\star} \right)^{1/3};\quad v_h = R_h \Omega,
  \label{eq:Rhill}
\end{equation}
with $a$ the disk radius (semi major axis), $M_\star$ the mass of the star, and $\Omega$ the orbital frequency at semi major axis $a$.  The escape velocity is defined $v_\mathrm{esc} = \sqrt{2GM/R}$ where $G$ is Newton's gravitational constant.  Using these definitions one can show that $v_\mathrm{esc}^2 = 6v_h^2/\alpha$,  where the dimensionless $\alpha$ is given by \citep[\cf][]{GoldreichEtal2004}
\begin{equation}
  \alpha \equiv \frac{R}{R_h} = 7.5\times10^{-3} \left( \frac{a}{\mathrm{AU}} \right)^{-1} \left( \frac{\rho}{\mathrm{g\ cm^{-3}}} \right)^{-\frac{1}{3}} \left( \frac{M_\star}{M_\odot} \right)^{\frac{1}{3}},
  \label{eq:alpha}
\end{equation}
where we used $M=4\pi \rho R^3/3$ with $\rho$ the internal density of the bodies.  When discussing interactions one should use combined masses and radii in the above definitions (\ie\ $M=M_1+M_2$ and $R=R_1+R_2$) but this leaves \eq{alpha} and the relation $v_\mathrm{esc}^2 = 6v_h^2/\alpha$ unaffected.

In the following we assume a two component model where a single big body of mass $M$ and radius $R$ interacts with smaller bodies of radius $R_0$ and mass $m_0$ that dominate the solids surface density, $\Sigma \approx \Sigma_{m0} \gg \Sigma_M$.  The velocity dispersion $v$ of the smaller bodies also dominates the relative velocity and we assume that $v_h < v < v_\mathrm{esc}$, where $v_h$ and $v_\mathrm{esc}$ correspond to, respectively, the Hill and escape velocity of the big body.  This is the dispersion dominated regime and these assumptions are typical for oligarchy. Using $\Sigma_M = M/(2\pi a \Delta a_\mathrm{st})$ with $\Delta a_\mathrm{st} =A R_h$ the width of the heating region and $A$ an order-of-unity factor, one finds that the runaway growth-oligarchy transition according to \eq{IM93} lies at a radius of (\cf\ \citealt{ThommesEtal2003})
\begin{eqnarray}
  \nonumber
  R_\mathrm{rg/oli} &=& \left[ \frac{3Aa\Sigma R_0^3}{4\rho \alpha} \right]^{1/5} = 
  94\ \mathrm{km} \left( \frac{A}{5} \right)^{1/5} \left( \frac{\rho}{\mathrm{g\ cm^{-3}}} \right)^{-2/15} \\
  & \times & \left( \frac{\Sigma}{10\ \mathrm{g\ cm^{-2}}} \right)^{1/5} \left( \frac{a}{\mathrm{AU}} \right)^{2/5} \left( \frac{R_0}{10\ \mathrm{km}} \right)^{3/5}.
  \label{eq:Roli}
\end{eqnarray}

In the dispersion-dominated regime inclinations $i$ are related to eccentricities $e$ as $i\approx e/2$ \citep{IdaEtal1993}.  The accretion rate of the big bodies then becomes
\begin{equation}
  \frac{dM}{dt} = C \pi \Sigma \frac{2GMR}{e^2 a^2 \Omega},
  \label{eq:dMdt}
\end{equation}
\citep{IdaMakino1993} where $e$ is the rms-eccentricity of the small bodies, which is related to $v$ as $v = ea\Omega$.  Furthermore, $C$ is a factor of order unity that takes into account the increased accretion rate under a distribution of velocities.  With the above definitions we transform \eq{dMdt} into Hill units
\begin{equation}
  \frac{dM}{dt} = \frac{6C\pi}{\alpha} R^2 \left( \frac{v_h}{v} \right)^2 \Sigma \Omega
\end{equation}
and define the accretion timescale in the 2-component approximation as
\begin{eqnarray}
  \nonumber
  T_\mathrm{ac} &=&\ \left( \frac{1}{M}\frac{dM}{dt} \right)^{-1} %= \left[ \frac{9C}{2} \left( \frac{v}{v_h} \right)^{-2} \frac{\Sigma \Omega}{\alpha \rho R} \right]^{-1} \\
  = \frac{2}{9C} \left( \frac{v}{v_h} \right)^2 \frac{\alpha R}{\Sigma/\rho} \Omega^{-1} \\
 &=&\ K_\mathrm{ac} \left( \frac{v}{v_h} \right)^2 \frac{\alpha \rho R}{\Sigma} \Omega^{-1},
 \label{eq:Tac}
\end{eqnarray}
where we defined the dimensionless $K_\mathrm{ac} = 2/9C$.

The timescale for viscous stirring of the small bodies is given by \citep{IdaMakino1993}:
\begin{equation}
  T_\mathrm{vs} = \frac{v^3}{4\pi G^2 n_M M^2 \ln \Lambda},% 
  \label{eq:Tvs1}
\end{equation}
where $\ln \Lambda$ is a Coulomb factor and $n_M$ the \textit{number density} of perturbers.  This latter quantity is obtained from the assumption that each small bodies `sees' one oligarch.  The volume traversed by the small bodies is the product of $2\pi a$ (their circumference), $2v/\Omega$ (the width in the radial direction), and $2v_z/\Omega$ (the vertical excursion). Here, $v_z\approx v/2$ represents the corresponding vertical velocities as given by the inclinations $i$ (=$e/2$) of the bodies. Thus, the effective number density of the single oligarch is $n_M = 1/(2\pi a) (2v/\Omega) (2v_z/\Omega)$.  Then, \eq{Tvs1} transforms into
\begin{eqnarray}
  \nonumber
  T_\mathrm{vs} &=& \frac{4}{\ln \Lambda} \frac{v^5 a}{v_\mathrm{esc}^4 R^2 \Omega^2}
  = \frac{1}{9 \ln \Lambda} \left( \frac{v}{v_h} \right)^5 \frac{a\alpha}{R} \Omega^{-1} \\
  &=& K_\mathrm{vs} \left( \frac{v}{v_h} \right)^5 \frac{a \alpha}{R} \Omega^{-1},
 \label{eq:Tvs}
\end{eqnarray}
with $K_\mathrm{vs} = 1/9\ln \Lambda $.

\section{The timescale for runaway growth, $T_\mathrm{rg}$}
\label{sec:Trg}
\begin{figure*}
  \includegraphics[width=0.45\textwidth]{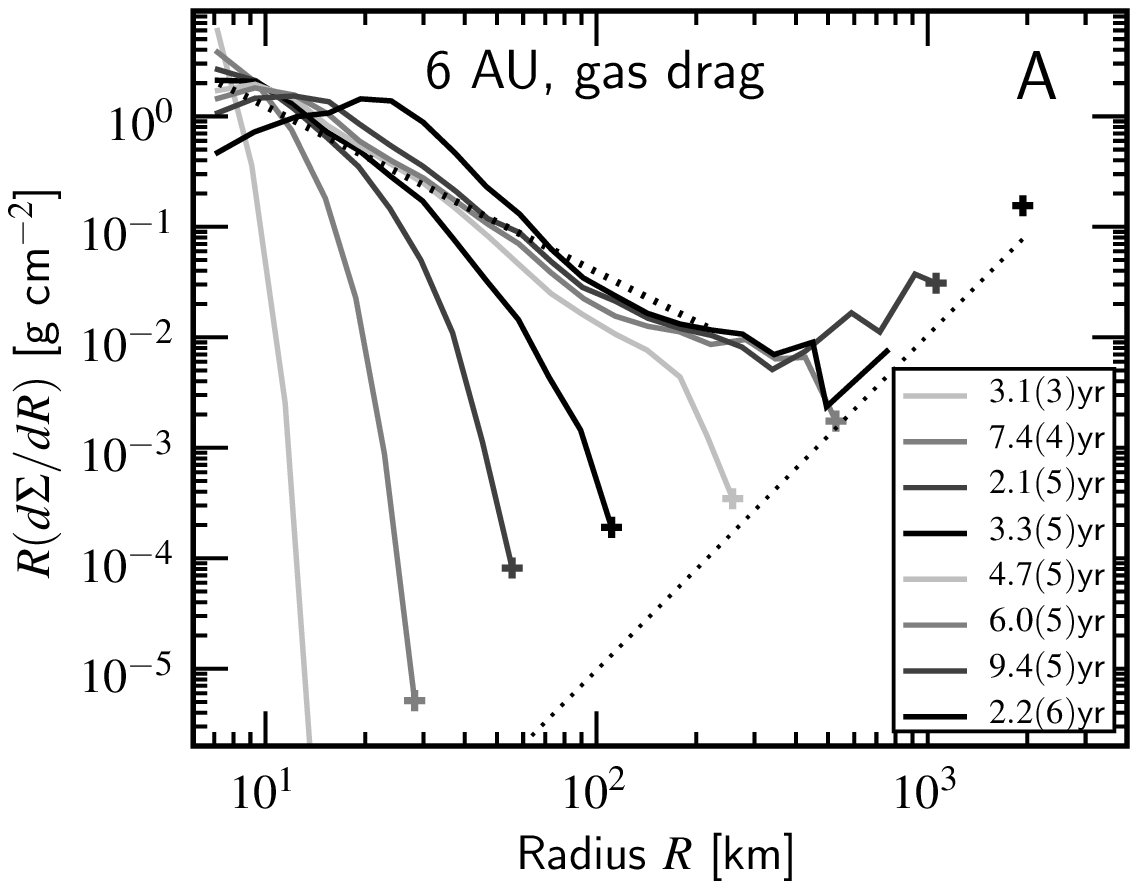}
  \includegraphics[width=0.45\textwidth]{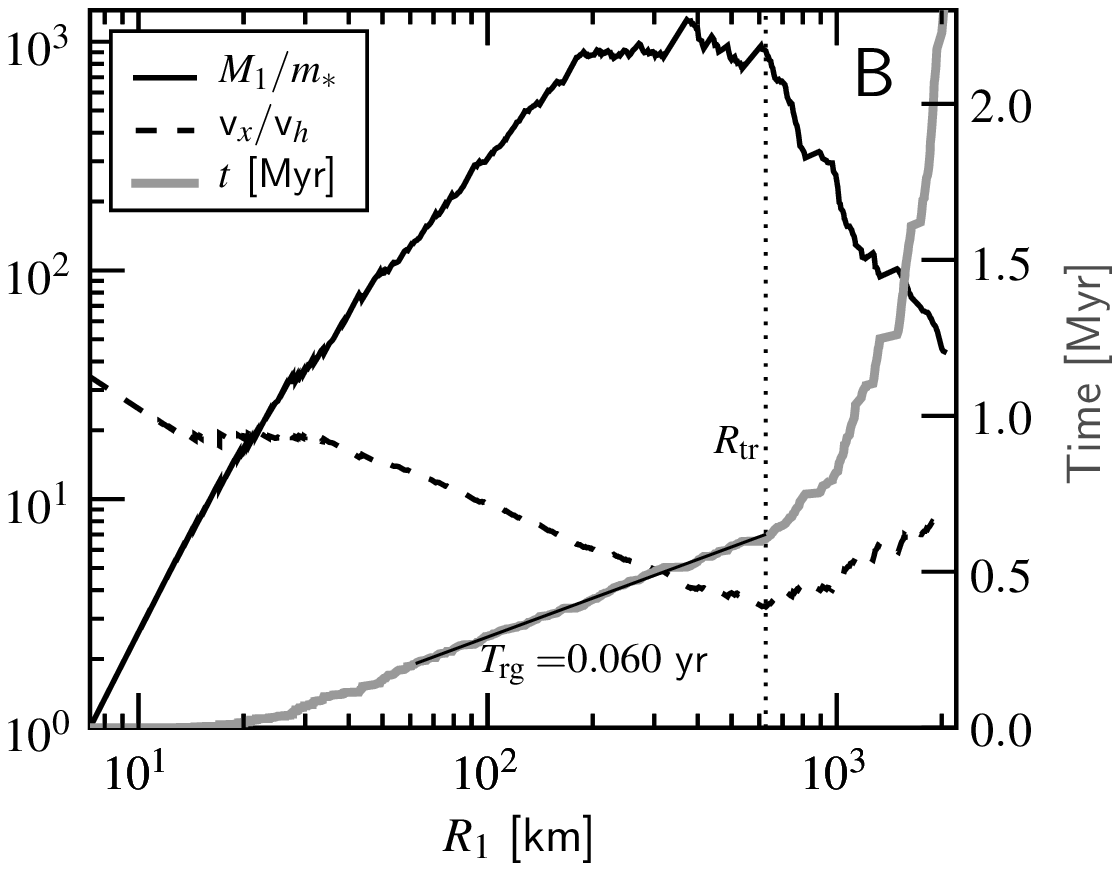}
  \caption{\label{fig:fig1} (\textit{left}) The mass spectrum $d\Sigma/dR$ for a simulation with gas drag conducted at 6 AU. After every factor of two increase of the radius of the most massive particle ($R_1$) the distribution is plotted.  The corresponding times are indicated in the legend. The \textit{dotted} line in the left corner indicates a $p=-2.5$ power-law slope for the column density spectrum, $N(m)$.  The dotted line in the bottom-right corresponds to a distribution that contains one body per bin. The simulation is continued until $R_1 = 2\times10^3\ \mathrm{km}$. (\textit{right}) Indicators of runaway growth: the mass of the most massive particle over the characteristic mass (\textit{solid black} curve) and the maximum velocity in the distribution ($v_x$) over the Hill velocity of the biggest body ($v_h$), which is a measure of the gravitational focusing (\textit{dashed black} curve).  These are plotted as function of the evolutionary parameter, $R_1(t)$.  Time is indicated by the \textit{grey curve} on the second y-axis.  The runaway growth timescale $T_\mathrm{rg}$ is obtained from the linear portion of the curve, before it steepens at $R_1 = R_\mathrm{tr}$.}
\end{figure*}

\Fg{fig1} presents an example of the runaway growth parameter study we have recently conducted \citep{OrmelEtal2010a}.  These are the results of statistical simulations that include key physical processes like dynamical friction, viscous stirring, gas drag, and resolve the semi-major axis of the bodies \citep[\cf][]{WeidenschillingEtal1997,BromleyKenyon2006}.  In \fg{fig1}a the mass spectrum is shown at several times during the simulation run.  After the radius of the most massive body, $R_1(t)$, has increased by a factor 2 a new curve is plotted; the corresponding times are indicated in the legend.   It is seen that the number density spectrum $N(m)$ evolves into a power law, $N(m)\propto m^{-p}$ with $p\approx-2.5$.  Near the end of the simulation bodies have separated: the oligarchs.

In \fg{fig1}b additional information for this run is presented as function of the evolutionary parameter $R_1(t)$.  The black curve shows the ratio of the most massive particle ($M_1$) and $m_\ast$, which is the mass-weighed average of the distribution, $m_\ast = \int m^2 N(m) dm /\int m N(m) dm$.  For a two component model $M_1/m_\ast$ should increase if the system is in runaway growth \citep{Lee2000,OrmelSpaans2008}.  It can be seen, however, that this ratio flattens near $R_1\sim200$ km and decreases after $\sim$600 km.  The fluctuations in this curve are caused by merging of bodies of similar size.  Also plotted is $v_x/v_h$, the ratio of the largest random velocity of the smallest bodies over the Hill velocity of $R_1$.  It first decreases (gravitational focusing increases) until $R_1\sim600$ km, where $v_x/v_h$ reaches a minimum.  This radius is denoted the transition radius, $R_\mathrm{tr}$, indicated by the vertical dotted line.  Time is plotted by the grey curve on the second, \textit{linear}, y-axis.  After $v_x/v_h$ has reached its minimum this curve noticeably steepens.  Only from this point onwards does dynamical heating win over the increase of $v_h$ through accretion, and the accretion timescale increases.  Bodies in other zones, spatially separated from $R_1$, then find the chance to catch up.

\begin{figure}
  \plotone{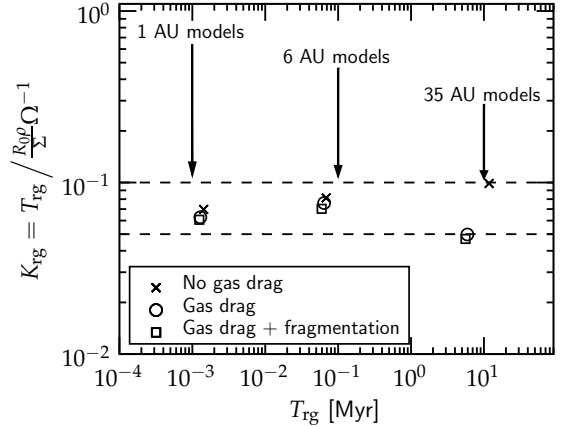}
  \caption{\label{fig:fig2}(x-axis) The timescale for accretion during runaway growth, $T_\mathrm{rg}$ for runaway growth simulations conducted at 1, 6, and 35 AU, and for models with and without gas drag and with (a simple prescription for) fragmentation. (y-axis) The dimensionless $K_\mathrm{rg}$, which gives $T_\mathrm{rg}$ normalized to the fiducial time $t_\mathrm{run} = R_0 \rho/\Sigma \Omega$, is much less than unity.  Runaway growth is fast.}
\end{figure}
The simulations show that during the runaway growth stage $M_1(t)$ grows exponentially with time ($R_1(t)$ is linear on a semilog plot) and an exponential fit is appropriate, $M_1(t) \propto \exp(t/T_\mathrm{rg})$ with $T_\mathrm{rg}$ the runaway growth timescale
(see \fg{fig1}b, thin solid line). In \fg{fig2} this quantity has been plotted for several runs performed at 1, 6, and 35 AU.  We express $T_\mathrm{rg}$ in terms of the fiducial timescale $t_\mathrm{run}=\rho R_0/\Sigma\Omega$ and derive the dimensionless $K_\mathrm{rg}$,
\begin{equation}
  T_\mathrm{rg} = K_\mathrm{rg} {t_\mathrm{run}} = K_\mathrm{rg} \frac{R_0 \rho}{\Omega \Sigma}.
  \label{eq:Trg}
\end{equation}
%In a two-component study performed by \citet{Rafikov2003ii}, $t_\mathrm{run}$ is the runaway growth timescale.   However, \fg{fig2} clearly shows that $T_\mathrm{rg} \ll t_\mathrm{run}$ ($K_\mathrm{rg}\ll 1$).  

During the initial evolution $T_\mathrm{rg} \ll T_\mathrm{ac}$ (\eqp{Tac}) and also $T_\mathrm{rg} \ll T_\mathrm{vs}$ (\eqp{Tvs}).    We find that the size and velocity spectrum that develops during runaway growth enables the biggest body to accrete particles from all masses, \ie\ also intermediate mass bodies \citep{OrmelEtal2010a}, which decreases $T_\mathrm{rg}$ compared to its 2-component estimate.  Similarly, despite the fact that random velocities ($v_x$) increase during runaway growth, the stirring by a \textit{single} biggest body is simply insignificant to shape the dynamical evolution of the smaller bodies.  Instead it is the \textit{ensemble} of large and intermediate mass bodies that produces the stirring.  A two component approximation is just too simple a picture for the runaway growth phase.  

\section{The runaway growth/Oligarchy transition}
\label{sec:rgot}
Although initially $T_\mathrm{vs} \gg T_\mathrm{rg}$ and $T_\mathrm{ac} \gg T_\mathrm{rg}$, during the runaway growth phase $v_x/v_h$ decreases and both $T_\mathrm{ac}$ and (especially) $T_\mathrm{vs}$ rapidly converge on $T_\mathrm{rg}$.  At the point where $T_\mathrm{vs}\lesssim T_\mathrm{rg}$ the runaway body starts to dynamically heat its environment at a rate faster than its previous accretion rate.  Thus, from this point the runaway growth can no longer `outpace' the stirring.  A single body dominates its neighborhood in terms of both velocity evolution and accretion, meaning a much slower growth rate because accretion now becomes self-regulated.  This is of course the key characteristic of oligarchy.  Conversely, we may suspect that the condition $T_\mathrm{vs} = T_\mathrm{rg}$ heralds the end of the runaway growth stage and the transition to oligarchy. 
\begin{figure}
  \plotone{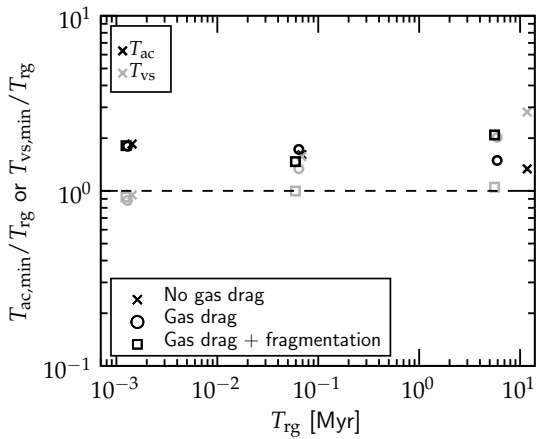}
  \caption{\label{fig:fig3}The accretion timescale for the two component assumption $T_\mathrm{ac}$ and $T_\mathrm{vs}$ (\eqps{Tac}{Tvs}) evaluated \textit{at the point where $v_x/v_h$ reaches its minimum value}, \ie\ at $R_1=R_\mathrm{tr}$ and $v_x/v_h= (v/v_h)_\mathrm{tr}$, normalized to $T_\mathrm{rg}$. }
\end{figure}

In \fg{fig3} we have plotted the ratio of these timescales over $T_\mathrm{rg}$ \textit{at the point where $v_x/v_h$ is at its minimum} (\ie\ at $R_1(t) = R_\mathrm{tr}$).  In evaluating $T_\mathrm{ac}$ and $T_\mathrm{vs}$ we have taken $C=3$ (\citealt{GreenzweigLissauer1992}; $K_\mathrm{ac} = 0.07$) and $\ln \Lambda = 3$ (\citealt{StewartIda2000}, $K_\mathrm{vs} = 0.04$). Because the $R_1$ \vs\ $v/v_h$ curve (\fg{fig1}b) differs from simulation to simulation (\eg\ when including gas drag and fragmentation) the timescales $T_\mathrm{ac}$ and $T_\mathrm{ev}$ also evolve differently.   However, from \fg{fig3} it is clear that at $R_1(t)=R_\mathrm{tr}$ all the relevant timescales are similar: $T_\mathrm{vs} \sim T_\mathrm{rg} \sim T_\mathrm{ac}$, irrespective of the simulation.  Thus, the 2-component model becomes first applicable at the point where gravitational focusing reaches its maximum.  This is the start of oligarchy.  From now on, $T_\mathrm{rg}$ loses its meaning (runaway growth ceases), $v_x/v_h$ increases, and accretion timescales are given by $T_\mathrm{ac}$ (which also increases).

Equating the relevant timescales, \ie\ \eqss{Tac}{Tvs}{Trg}, we solve for both $R_\mathrm{tr}$ and $(v_x/v_h)_\mathrm{tr}$  %Identifying the former with the runaway growth-oligarchy transition, $R_\mathrm{tr}$, we find 
\begin{eqnarray}
  \label{eq:R-rg1}
  R_\mathrm{tr} &=& \left[ \frac{K_\mathrm{vs}^2 K_\mathrm{rg}^3}{K_\mathrm{ac}^5} \left( \frac{a\Sigma}{\rho} \right)^2 \left( \frac{R_0}{\alpha} \right)^3 \right]^{1/7},\\ 
  \label{eq:uvhtr}
  \left( \frac{v_x}{v_h} \right)_\mathrm{tr} &=& \sqrt{\frac{K_\mathrm{rg} R_0}{\alpha K_\mathrm{ac}R_\mathrm{tr}}}.
\end{eqnarray}
\Eq{R-rg1} supersedes $R_\mathrm{rg/oli}$ of \eq{Roli} as the new criterion between the runaway growth and oligarchy accretion phases.  Using the above expressions for $K_\mathrm{ac}$ and $K_\mathrm{vs}$ and inserting \eq{alpha} for $\alpha$ we find that \eq{R-rg1} transforms into
\begin{eqnarray}
  \nonumber
  R_\mathrm{tr} &=& 320\ \mathrm{km} \left( \frac{K_\mathrm{rg}}{0.1} \right)^{3/7} \left( \frac{\rho}{1\ \mathrm{g\ cm^{-3}}} \right)^{-1/7} \\
  & \times & \left( \frac{R_0}{10\ \mathrm{km}} \right)^{3/7} \left( \frac{a}{\mathrm{AU}} \right)^{5/7} \left( \frac{\Sigma}{10\ \mathrm{g\ cm^{-2}}} \right)^{2/7},
  \label{eq:R-rg}
\end{eqnarray}
where the prefactor is $\sim$3 times larger than \eq{Roli}.  Any order-of-unity uncertainty in the relation $T_\mathrm{rg}=T_\mathrm{ac}=T_\mathrm{vs}$ can be regarded as an uncertainty in the $K$-factors, but this affects the prefactor only marginally.  More fundamentally, \eq{R-rg} shows different dependencies on $R_0$, $\Sigma$, and, especially, $a$.  If $R_0$ is large, \eg\ $R_0\sim 500$ km, $R_\mathrm{rg/oli}$ and $R_\mathrm{tr}$ give approximately the same values at 1 AU.  A large $R_0$ is the favored outcome of recent numerical simulations involving a turbulent layer of densely packed boulders \citep{JohansenEtal2007,JohansenEtal2009}.  Then, the domain of runaway growth is rather limited (a factor three in size); but it still produces $\sim$10$^3$ km objects.

\begin{figure}
  \plotone{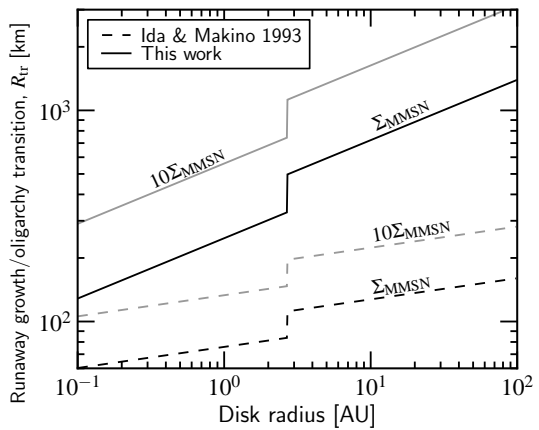}
  \caption{\label{fig:fig4}The runaway/oligarchy transition, $R_\mathrm{tr}$, against disk radius according to \eq{Roli} (\citealt{IdaMakino1993}, dashed lines) and \eq{R-rg} (this work, \textit{solid} lines). Plotted are $R_\mathrm{tr}$ for a minimum mass solar nebula (MMSN) profile of surface densities (\textit{black} lines) and a 10 times enhancement of $\Sigma$ over the MMSN value (\textit{grey} lines).  The MMSN profile is adopted from \citet{NakagawaEtal1986}.}
\end{figure}
In \fg{fig4} we illustrate the runaway growth/oligarchy turnover point of \citet{IdaMakino1993} (\eqp{Roli}, \textit{dashed} lines) and \eq{R-rg} (\textit{solid} lines) as function of distance $a$ for an initial planetesimal radius of $R_0=10\ \mathrm{km}$ and an internal density of $\rho=3\ \mathrm{g\ cm^{-3}}$.  For $\Sigma(a)$ we adopt a Minimum Mass Solar Nebula (MMSN) profile \citep{Weidenschilling1977} (\textit{black} lines) and a profile ten times larger than this (\textit{grey} lines).  In \Fg{fig4} it can be seen that the difference between the two criteria becomes largest in the outer disk ($a \gg 10$ AU):  whereas according to the \citet{IdaMakino1993} prescription the transition takes place at radii of $\sim$100 km, the new criterion shifts it upwards to $\sim$10$^3$ km.  

\section{Discussion}
\label{sec:summ}
According to our findings, the transition between the runaway growth and oligarchic growth phases is characterized by the following properties:
\begin{itemize}
  \item A power-law size distribution of mass index $p\approx-2.5$, extending from the initial size $R_0$ to the transition size $R_\mathrm{tr}$ as given by \eq{R-rg};
  \item The random velocity $v$ of the planetesimal bodies (of size $\sim$$R_0$), via \eq{uvhtr};
  \item The timescale, via \eq{Trg}.  Here, $T_\mathrm{rg}$ must be multiplied by a term $\sim$$\log (R_\mathrm{tr}/R_0)$ to account for the several e-foldings of enjoyed growth.  Then we obtain $t=T_\mathrm{tr} \sim \rho R_0/\Sigma_0 \Omega$ for the time until the transition.  It is remarkable that this short timescale depends on the initial conditions only, a result that is unique to the runaway growth phase.
\end{itemize}
We assess the implications of these findings for the broader context of planet formation.  First, we do not expect the final timescales of core formation to be much influenced by the new transition radius, since these are set by the much slower oligarchy stage that supersedes runaway growth.  Indeed, semi-analytical studies of oligarchic growth required additional mechanisms like gas damping, fragmentation, or migration to produce embryos on reasonable timescales \citep[\eg][]{BruniniBenvenuto2008,Chambers2008}.  On the other hand, gap formation \citep{Rafikov2003ii} will increase formation timescales.   

Recently, \citet{LevisonEtal2010} investigate core-formation scenarios using $N$-body techniques.  The aim of that study was to understand how efficient planetesimal accretion proceeds during the phase where an Earth-size planetary embryo has to grow to a mass of $\sim$$10M_\oplus$ to be able to accrete the nebula gas and become a gas giant.  They recognize the importance of processes like planetesimal scattering, planetesimal orbital decay due to gas drag (both processes quench the growth), and planetesimal-driven embryo migration (which is conducive to growth).  The outcome of these simulations, furthermore, is found to depend on the initial setup of the simulation, \ie\ the size distribution of the planetesimals. Given the significance that is attached to the wholesale redistribution of matter, it would also be of interest to assess the importance of these effects for the early oligarchy stage, \eg\ to perform $N-$body simulations with embryos radii starting at the transition mass $R_\mathrm{tr}$.   

The second implication of our study concerns the Kuiper belt. Our results strongly suggest that the Kuiper Belt is primarily the product of the runaway growth phase.  First, assuming that the initial surface density is approximately MMSN or larger ($\Sigma \sim 0.1\ \mathrm{g\ cm^{-2}}$), the biggest $\sim$10$^3$ km bodies (plutinos) can be produced by runaway growth.  Second, the observed mass distribution $N(m)$ for the largest Kuiper belt objects (KBOs) obeys a power-law with $p\approx-2.5$.  Recent studies find a power-law size index $q$ (as in $N(R) \propto R^{-q}$) of $4.8\pm0.3$ ($p=-2.3\pm0.1$; \citealt{FraserKavelaars2009}) and $4.5^{+1.0}_{-0.5}$ ($p=-2.2^{+0.2}_{-0.3}$; \citealt{FuentesHolman2008}). (The size distribution of the Kuiper belt's smallest bodies is collisionally dominated, though, and $q$ is much lower.)  The large $q$, together with the fact that the KBO size distribution is continuous rather than bimodal as in the end stage of \fg{fig1}a, argue that it has evolved only very little since runaway growth and has not been significantly shaped by oligarchic growth.  

In order to produce the KBOs in a sufficiently short time span, Kuiper Belt formation scenarios \citep[\eg][]{KenyonLuu1998,ChiangEtal2007} assume that the initial belt contained much more mass than the $\sim0.01\ M_\oplus$ ($\Sigma \sim 0.001\ \mathrm{g\ cm^{-2}}$) that is present today \citep{BernsteinEtal2004}.  Neptune formation and/or another dynamical shakeup event (as in the Nice model) subsequently depleted 99\% of its mass \citep{FordChiang2007,LevisonEtal2008}.  Using our results, we can verify these findings.  Assuming $a=35$ AU, we find
\begin{eqnarray}
  \label{eq:RT-kuipera}
  R_\mathrm{tr} &\sim& 10^3\ \mathrm{km} \left( \frac{R_0}{\mathrm{10\ km}} \right)^{3/7} \left( \frac{\Sigma}{\mathrm{0.1\ g\ cm^{-2}}} \right)^{2/7};  \\
  \label{eq:RT-kuiperb}
  T_\mathrm{tr} &\sim& \frac{R_0 \rho}{\Omega \Sigma} \sim 10^8\ \mathrm{yr}\left( \frac{R_0}{\mathrm{10\ km}} \right) \left( \frac{\Sigma}{\mathrm{0.1\ g\ cm^{-2}}} \right)^{-1},
\end{eqnarray}
%Fixing $R_\mathrm{tr}\sim10^3$ km in \eq{RT-kuipera} we find a relation between $R_0$ and $\Sigma$.  Inserting this into \eq{RT-kuiperb} we obtain $T_\mathrm{tr} \sim 10^8\ \mathrm{yr}\ (\Sigma/\mathrm{0.1\ g\ cm^{-2}})^{-5/3}$, 
which readily shows the need for an enhanced surface density over the current one: for $\Sigma \sim 0.001\ \mathrm{g\ cm^{-2}}$ either $R_\mathrm{tr}$ becomes too low or $T_\mathrm{tr}$ too long.  Thus, we conclude that the KBO size distribution as seen today is consistent with a scenario of being a leftover product of the initial runaway growth phase and has since been depleted.  These findings are in line with the simulations of \citet{KenyonBromley2008,KenyonBromley2009arXiv}, where growth also stalls after $\sim$$10^3$ km bodies have been formed.  Further growth is impeded since oligarchic accretion timescales become too long, even at enhanced surface densities.

\acknowledgements
We acknowledge the helpful comments of the (anonymous) referee and his/her encouragement to put our results into a broader context.  C.W.O.\ is supported by a grant from the Alexander von Humboldt foundation.

\end{document}